\documentstyle[12pt,epsfig]{article}
\def\be{\begin{equation}}
\def\ee{\end{equation}}
\def\bc{\begin{center}}
\def\ec{\end{center}}
\def\bea{\begin{eqnarray}}
\def\eea{\end{eqnarray}}

\def\nn{\nonumber}

\def\cm{{\cal M}}
\def\co{{\cal O}}
\def\ctmax{{|\cos \theta_{\gamma}|_{max}}}
\def\ctg{\cos \theta_{\gamma}}
\def\dslash{\not{\! \partial}}
\def\Dslash{\not{\!\! D}}
\def\emu{\epsilon_{\mu}}
\def\esl{\not{\! \epsilon}}
\def\ev{{\rm \; eV}}
\def\gev{{\rm \; GeV}}
\def\gmd{\gamma_{\mu}}
\def\gmu{\gamma^{\mu}}
\def\gc{\gamma_5}
\def\grav{\tilde{G}}

\def\ksl{\not{\! k}}
\def\mfre{{ {\rm Re} \left( {M \over F} \right)}}
\def\mfim{{ {\rm Im} \left( {M \over F} \right)}}
\def\mpl{M_{\rm P}}
\def\ov{\overline}
\def\pdsl{\not{\! p}_2}
\def\psie{e}
\def\psieb{\overline{e}}

\def\psis{\tilde{G}}
\def\psisb{\overline{\tilde{G}}}
\def\psil{\lambda}
\def\psilb{\overline{\lambda}}
\def\te{\tilde{e}}
\def\tec{\tilde{e^{c}}}
\def\tg{\tilde{G}}
\def\dmu{\partial^\mu}
\def\dmd{\partial_\mu}
\def\Dmu{D^\mu}
\def\Dmd{D_\mu}
\def\pusl{\not{\! p}_1}
\def\qdsl{\not{\! q}_2}
\def\qusl{\not{\! q}_1}

\def\simgt{\stackrel{>}{{}_\sim}}
\def\sqtg{\sin^2 \theta_{\gamma}}

\def\usbu{\overline{u}_G(q_1)}
\def\usbd{\overline{u}_G(q_2)}
\def\upu{u(p_1)}
\def\vbd{\overline{v}(p_2)}
\def\vsu{v_G(q_1)}
\def\vsd{v_G(q_2)}
\def\xg{x_{\gamma}}
\def\xgmin{{x_{\gamma,min}}}

\catcode`@=11
\def\marginnote#1{}
\newcount\hour
\newcount\minute
\newtoks\amorpm
\hour=\time\divide\hour by60
\minute=\time{\multiply\hour by60 \global\advance\minute by-\hour}
\edef\standardtime{{\ifnum\hour<12 \global\amorpm={am}%
        \else\global\amorpm={pm}\advance\hour by-12 \fi
        \ifnum\hour=0 \hour=12 \fi
        \number\hour:\ifnum\minute<10 0\fi\number\minute\the\amorpm}}
\edef\militarytime{\number\hour:\ifnum\minute<10 0\fi\number\minute}
\def\draftlabel#1{{\@bsphack\if@filesw {\let\thepage\relax
   \xdef\@gtempa{\write\@auxout{\string
      \newlabel{#1}{{\@currentlabel}{\thepage}}}}}\@gtempa
   \if@nobreak \ifvmode\nobreak\fi\fi\fi\@esphack}
        \gdef\@eqnlabel{#1}}
\def\@eqnlabel{}
\def\@vacuum{}
\def\draftmarginnote#1{\marginpar{\raggedright\scriptsize\tt#1}}
\def\draft{\oddsidemargin 0.0truein
        \def\@oddfoot{\sl preliminary draft \hfil
        \rm\thepage\hfil\sl\today\quad\militarytime}
        \let\@evenfoot\@oddfoot \overfullrule 3pt
        \let\label=\draftlabel
        \let\marginnote=\draftmarginnote
   \def\@eqnnum{(\theequation)\rlap{\kern\marginparsep\tt\@eqnlabel}%
\global\let\@eqnlabel\@vacuum}  }
\catcode`@=12
\begin{document}
\begin{titlepage}
\vspace*{-1cm}
hep-ph/9711516
\hfill{CERN-TH/97-339}
\\
\phantom{bla}
\hfill{DFPD~97/TH/52}
\\
\vskip 1.0cm
\begin{center}
{\Large \bf Signals of a superlight gravitino at
$e^+ e^-$ colliders
\\
when the other superparticles are heavy\footnote{Work
supported in part by the European Commission TMR
Programme ERBFMRX-CT96-0045.}}
\end{center}
\vskip 0.8  cm
\begin{center}
{\large Andrea Brignole}\footnote{e-mail address:
brignole@vxcern.cern.ch},
{\large Ferruccio Feruglio}\footnote{e-mail address:
feruglio@padova.infn.it}$^,$\footnote{On leave from
Dipartimento di Fisica, Universit\`a di Padova,
I-35131 Padua, Italy}
\\
\vskip .1cm
Theory Division, CERN, CH-1211 Geneva 23, Switzerland
\\
\vskip .1cm
and
\\
\vskip .1cm
{\large Fabio
Zwirner}\footnote{e-mail address: zwirner@padova.infn.it}$^
,$\footnote{Also at Dipartimento di Fisica, Universit\`a 
di Padova, I-35131 Padua, Italy}
\\
\vskip .1cm
INFN, Sezione di Padova, I-35131 Padua, Italy
\end{center}
\vskip 0.5cm
\begin{abstract}
\noindent
If the gravitino $\grav$ is very light and all the other
supersymmetric particles are above threshold, supersymmetry
may still be found at colliders, by looking at processes with
only gravitinos and ordinary particles in the final state.
We compute here the cross-section for the process $e^+ e^-
\to \grav \grav \gamma$, whose final state can give rise to 
a distinctive photon plus missing energy signal at present 
and future $e^+ e^-$ colliders. We describe how the present
LEP data can be used to establish a lower bound on the
gravitino mass of order $10^{-5}$~eV. We conclude with a 
critical discussion of our results, comparing them with 
related ones and outlining possible generalizations.
\end{abstract}
\vfill{
CERN-TH/97-339
\newline
\noindent
November 1997}
\end{titlepage}
\setcounter{footnote}{0}
\vskip2truecm
\section{Introduction}

Low-energy supersymmetry (for reviews and references,
see e.g. \cite{susy}) is the most popular and probably
the best motivated extension of the Standard Model, at
energy scales accessible to present or future colliders.
At a fundamental level, however, the present dynamical
understanding of supersymmetry breaking is still quite
unsatisfactory. Denoting by $\Delta m$ the scale of the
supersymmetry-breaking mass splittings between the
Standard Model particles and their superpartners, and
by $\Lambda_S$ the microscopic supersymmetry-breaking
scale (in one-to-one correspondence with the gravitino
mass $m_{3/2}$), from a phenomenological viewpoint both
these scales should be taken as arbitrary parameters.
Different possibilities then arise, as reviewed in a
recent paper of ours \cite{bfz3}, to which we refer for
more details on the general theoretical framework: we
denoted these possibilities by `heavy', `light' and
`superlight' gravitino. In the heavy gravitino case,
reactions involving the gravitino are never important for
collider physics. In the light gravitino case, the
gravitino can be relevant in the decays of other 
supersymmetric particles, if there is sufficient energy 
to produce the latter. In the superlight gravitino
case, also the direct production of gravitinos (with
or without other supersymmetric particles) can become
relevant. In this paper, we concentrate on the superlight 
gravitino case, where the two scales $\Delta m$ and $\Lambda_S$ 
are both close to the electroweak scale and to each other, 
the gravitino mass $m_{3/2}$ is several orders of magnitude 
below the $\ev$ scale and, as will be clear soon, the resulting 
phenomenology can be quite different from the extensively 
studied cases of heavy and light gravitino.

Many aspects of the superlight gravitino phenomenology at
colliders have been discussed long ago in the pioneering
works by Fayet \cite{fayet,pierre}, and also in more recent
papers \cite{slgcol}. All these authors, however, assumed
that some other supersymmetric particle, for example a
neutralino or one of the spin-0 partners of the gravitino,
is light enough to be produced on-shell in some reaction.
Here we take an orthogonal point of view: there may be
experiments where the available energy is still
insufficient for the on-shell production of other
supersymmetric particles, but nevertheless sufficient to
give rise to final states with only gravitinos and ordinary
particles, at measurable rates.

As discussed in \cite{bfz3}, particularly powerful processes to
search for a superlight gravitino $\grav$ (when the supersymmetric
partners of the Standard Model particles and the spin--0
superpartners of the gravitino are above threshold) are:
\be
e^+ e^- \longrightarrow \grav \grav \gamma \, ,
\label{epem}
\ee
and $q \ov{q} \longrightarrow \grav \grav \gamma$, which 
would give rise to a distinctive $(photon + missing
\; energy)$ signal. The first process can be studied at
$e^+ e^-$ colliders such as LEP or the proposed NLC, the
second one at hadron colliders such as the Tevatron or
the LHC. At hadron colliders, we can also consider the
process $q \ov{q} \longrightarrow \grav \grav g$, which, 
together with other partonic subprocesses such as $q g
\to q \grav \grav$ and $g g \to g \grav \grav$, contributes 
to the $(jet + missing \; energy)$ signal. In this letter, we 
compute the cross-section and the relevant angular distributions 
for the process of eq.~(\ref{epem}), in the limit in which the
supersymmetric particles of the Minimal Supersymmetric extension
of the Standard Model (MSSM), such as sleptons, squarks and gauginos,
and all other exotic particles, such as the spin-0 partners of the
gravitino, are heavy. We also show how the present LEP data, in 
the absence of a signal over the Standard Model background, 
should allow to establish the lower bound $\Lambda_S \simgt
200 \gev$, or, equivalently, $m_{3/2} \simgt 10^{-5} \ev$.
In contrast with other collider bounds discussed in the 
literature \cite{slgcol}, the ones discussed in the present
paper cannot be evaded by modifying the mass spectrum of the 
other supersymmetric particles: making some additional 
supersymmetric particle light leads in general to stronger 
bounds. The study of the processes involving quarks and/or 
gluons and the discussion of the phenomenology at hadron 
colliders will be presented in a companion paper \cite{bfmz}.

For the theoretically oriented readers, we summarize the
framework of our calculation. The $\pm 1/2$ helicity
components of the superlight but massive gravitino,
corresponding to the would-be goldstino, have
effective couplings to the MSSM fields comparable in
strength with the MSSM gauge couplings. Exploiting the
supersymmetric version of the equivalence theorem
\cite{equiv}, we start from an effective theory where global
supersymmetry is linearly realized but spontaneously broken.
Its degrees of freedom are just the MSSM superfields and
the superfields containing the goldstino $\grav$ (denoted here,
in the spirit of the equivalence theorem, with the same symbol
as the gravitino). Assuming pure $F$-breaking, as we will do 
in the following, it is sufficient to add to the MSSM states 
a single neutral chiral superfield. Such an effective theory 
is non-renormalizable, and its Lagrangian can be parametrized  
(up to higher-derivative and Fayet-Iliopoulos terms, assumed 
here to be negligible) by a K\"ahler potential $K$, a 
superpotential $w$ and a gauge kinetic function $f$. However, 
the terms that contribute to the on-shell scattering amplitudes 
for the process of eq.~(\ref{epem}) do not depend on
the details of these defining functions, but only
on suitable combinations of $\Delta m$ and $\Lambda_S$.
We then move to a `more effective' theory by explicit integration
of the heavy superpartners in the low-energy limit. The
residual degrees of freedom are only the goldstino and the
Standard Model particles, and at this level supersymmetry
is non-linearly realized \cite{nl} in a non-standard way
\cite{bfzeegg}. We finally compute the relevant scattering
amplitudes and the cross-section. As we shall see, the final 
result is independent of $\Delta m$, and is function only of 
the centre-of-mass energy $\sqrt{s}$ and of the 
supersymmetry-breaking scale $\Lambda_S$ (or, equivalently, 
of the gravitino mass $m_{3/2}$).

The paper is organized as follows. In sect.~2 we define in full
detail the theoretical framework of our calculation, identifying
the explicit form of the relevant lagrangian terms, at the level
where supersymmetry is linearly realized but spontaneously broken.
In sect.~3 we show how to take the low-energy limit in the case 
of heavy superpartners, ending up with an effective lagrangian
where supersymmetry is non-linearly realized. We also list the 
independent amplitudes contributing, in this limit, to the process 
under consideration. In sect.~4 we compute the differential and the
integrated cross-section, and discuss the implications of the 
present LEP data. Finally, in sect.~5 we give a critical discussion
of our results and outline further generalizations.  

Before concluding this introduction, we should mention
some related literature. The process of eq.~(\ref{epem}) was
already considered by Fayet \cite{pierre}, who gave the scaling
laws of the cross-section  with respect to the gravitino mass and
the centre-of-mass energy in the low-energy limit. However, at the
time the possibility of heavy gauginos was not considered, and, 
in the limit of light photino, other processes with photinos in 
the final state become more important. The process of
eq.~(\ref{epem}) was also considered in
\cite{nach}, in the same kinematical limit as in the present
work, but relying on the non-linear realization of supersymmetry
proposed in \cite{wess}. As will be discussed in the
concluding section, the two treatments are not equivalent and
lead to different results, both consistent with supersymmetry.
The effective interactions of a
superlight gravitino with the Standard Model fermions and gauge
bosons have been recently discussed in \cite{lp}, which identified
possible new effects associated with Fayet-Iliopoulos terms: these
effects do not arise in our calculational framework.

\section{Theoretical framework}

To discuss the process of eq.~(\ref{epem}),
given the present collider energies and the typical
cuts on the photon spectrum, we can safely neglect
the electron mass. Our starting point is an $N=1$ globally 
supersymmetric theory with gauge group $U(1)_{em}$. Besides
the vector supermultiplet, describing the photon $A_\mu$ and the
photino $\lambda$, the model contains three chiral superfields,
with spinor and scalar components $\psi_i$ and $\varphi_i$,
respectively $(i=0,1,2)$. Two of them $(i=1,2)$ describe the
Dirac electron field $\psie$ together with the selectrons
(complex scalar fields) $\te \equiv \te_L$ and $\tec \equiv
(\te_R)^*$. The remaining chiral multiplet, $(i=0)$, is
electrically neutral, and describes a Majorana fermion, $\tg$,
and a complex scalar $z \equiv (S + i P) / \sqrt{2}$. The
theory is specified, up to higher-derivative and Fayet-Iliopoulos
terms, to be neglected here, in terms of three defining functions: 
the superpotential $w$, the K\"ahler potential $K$ and the gauge 
kinetic function $f$. The explicit form of the lagrangian in terms 
of $w$, $K$ and $f$ is standard: we refer to the Appendix of 
ref.~\cite{bfz3} for its expression in component fields. We assume 
that the defining functions of the theory are such that, at the 
minimum of the scalar potential,
\be
\langle  F^0  \rangle \ne 0 \, ,
\;\;\;\;\;
\langle  F^{1,2}  \rangle = 0 \, ,
\label{fterms}
\ee
where $F^i$ denote the auxiliary fields of the chiral
supermultiplets. Therefore, supersymmetry is spontaneously
broken and the goldstino coincides with $\tg$. We will also
assume $\langle \varphi^i \rangle \! =0$ $(i=1,2)$, so that
$U(1)_{em}$ remains an exact local symmetry.

As anticipated, we consider the goldstino interactions in the spirit
of the equivalence theorem. The scattering amplitudes involving
external
goldstinos are identified with corresponding scattering amplitudes
with external gravitinos of an $N=1$ locally supersymmetric theory.
This equality holds up to terms that, at fixed scattering angles,
are of order $m_{3/2}/\sqrt{s}$, where $\sqrt{s}$ is the
centre-of-mass energy of the process. As we shall see, we are always
going to consider situations where such correction terms are
harmless: typically, $\sqrt{s}$ will be more than ${{\cal O}} (100 
\gev)$, whilst we will be sensitive to values of $m_{3/2}$ less 
than ${{\cal O}} (10^{-4} \ev)$.

We proceed by expanding the defining functions of the theory
around the vacuum, in order to identify the terms potentially
relevant for the process under consideration. To carry
on such an expansion in a simple way, we characterize the
functional dependence of $w$, $K$ and $f$ as follows:
\bea
w&=&\hat{w}(z)+ \ldots \, ,
\nn\\
K&=&\hat{K}(z,\bar{z})+\tilde{K}(z,\bar{z})(|\te|^2+|\tec|^2)
+\ldots \, ,
\nn\\
f&=&\hat{f}(z)+\ldots \, ,
\label{expan}
\eea
where the dots denote terms which will not take part in our
discussion. For simplicity, in the K\"ahler potential of  
eq.~(\ref{expan}) the bilinears in the selectron fields 
have been chosen to be equal: this will give rise to common 
supersymmetry-breaking masses for the two selectrons. The
reader can easily check that a non-universal choice for the
coefficients of $|\te|^2$ and $|\tec|^2$ in $K$, giving rise
to different supersymmetry-breaking masses for $\te$ and $\tec$,
would not affect the low-energy behaviour of the amplitudes 
considered here. Also, following the standard practice,
we have assumed negligible mixing in the selectron sector. 
Taking into account eqs.~(\ref{fterms}) and (\ref{expan}),
it is straightforward to evaluate the mass spectrum of the model 
and the relevant interactions, along the lines of \cite{bfz3}.
The photon, the goldstino and the electron remain massless,
whereas all the other particles acquire non-vanishing masses
proportional to $\langle F^0 \rangle$ and expressed in terms of
$w$, $K$, $f$ and their derivatives, evaluated on the vacuum.
Moreover, even in the presence of non-renormalizable interactions,
the expansion of the lagrangian in (canonically normalized)
component fields can be rearranged in such a way that all the
terms of interest for our calculation are expressed in terms of
the mass parameters $(m_{S,P}^2,m^2,M)$, associated with the
spin-0 partners of the goldstino, the selectrons and the
photino respectively, and the scale $F$ of supersymmetry
breaking, without explicit reference to $w$, $K$ and $f$:
\bea
{\cal L}
& = &
-{1\over 4} F_{\mu\nu} F^{\mu\nu}
+{i\over 2} \psilb\dslash\psil +{1\over 2}|M|~\psilb\psil
+{i\over 2} \psisb\dslash\psis
\nn\\
&+&{1\over 2} (\dmu S) (\dmd S) -{1\over 2} m_S^2 S^2
+ {1\over 2} (\dmu P) (\dmd P) -{1\over 2} m_P^2 P^2
\nn\\
&+& i\psieb\Dslash\psie
+(\Dmu \te_L)^* (\Dmd\te_L)-m^2|\te_L|^2
+(\Dmu \te_R)^* (\Dmd\te_R)-m^2|\te_R|^2
\nn\\
&-&\sqrt{2}i~ g~
\left(\te_L^*~\psilb P_L\psie-\te_L~\psieb P_R\psil
-\te_R~\psieb P_L\psil+\te_R^*~\psilb P_R\psie\right)
\nn\\
&-&{1\over 2\sqrt{2}} {m_S^2\over |M|}
S\left[ \mfre \psisb\psis-i \mfim \psisb\gc\psis\right]
\nn\\
&-&{1\over 2\sqrt{2}} {m_P^2\over |M|}
P\left[ \mfim \psisb\psis+i \mfre \psisb\gc\psis\right]
\nn\\
&-& {1\over 2\sqrt{2}}\left[ \mfre S- \mfim P\right]
F_{\mu\nu} F^{\mu\nu}
+ {1\over 2\sqrt{2}}\left[ \mfim S+ \mfre P\right]
F_{\mu\nu}\tilde{F}^{\mu\nu}
\nn\\
&-&{1\over 4\sqrt{2}} \psisb [\gamma^\mu,\gamma^\nu]\psil
\left[ \mfre F_{\mu\nu} - \mfim \tilde{F}_{\mu \nu} \right]
\nn\\
&-&{m^2\over |M|} {M \over F} \left(\te_L^*~ \psisb P_L \psie +
\te_R~\psieb P_L \psis \right)
-{m^2\over |M|} {M^* \over F^*} \left(\te_R^* \psisb P_R \psie +
\te_L~\psieb P_R \psis \right)
\nn\\
&-&{m^2\over 2 |F|^2} \left(\psisb\psie~ \psieb\psis-
\psisb\gc\psie~ \psieb\gc\psis\right)
+ \ldots \, .
\label{lag}
\eea
In eq.~(\ref{lag}), $P_L=(1-\gc)/2$, $P_R=(1+\gc)/2$ and
$\tilde{F}_{\mu\nu}= 1/2~\epsilon_{\mu\nu\rho\sigma}
F^{\rho\sigma}$, where $\epsilon_{0123} = - \epsilon^{0123}
= 1$. The parameter $F=<\ov{w}_{\ov{z}}
(K_{\ov{z} z})^{-1/2}>$ (lower indices denote here
derivatives) has the dimension of a mass squared and 
defines the supersymmetry-breaking scale, $\Lambda_S = 
|F|^{1/2}$. We remind the reader that, in our flat space-time, 
$|F|$ is linked to the gravitino mass $m_{3/2}$ by the 
universal relation
\be
\left| F \right| = \sqrt{3} \, m_{3/2} \, \mpl \, ,
\;\;\;\;\;
\left[ \mpl \equiv (8 \pi G_N)^{-1/2} \simeq
2.4 \times 10^{18} \gev \right] \, .
\label{flat}
\ee
The phases of $M$ and $F$ are assumed to be arbitrary. By means
of field redefinitions, we can remove all the phases but one,
assigned here to the ratio  $M/F$. For fermions, we have used
a four-component notation. The covariant derivatives are given by
\be
\Dmd e= \left(\dmd - i~g~ A_\mu\right) e,~~~~~
\Dmd \te= \left(\dmd - i~g~ A_\mu\right) \te,~~~~~\Dmd \tec
= \left(\dmd + i~g~ A_\mu\right) \tec \, ,
\label{cov}
\ee
where $g$ is the coupling constant of $U(1)_{em}$.
Finally, the dots in eq.~(\ref{lag}) stand for terms that do
not contribute to the amplitude evaluated here\footnote{There
are interaction terms proportional to $<\tilde{K}_S>$ and
$<\tilde{K}_{\ov{S}}>$, not explicitly listed here, that are
in principle relevant. An explicit computation shows that their
total contribution vanishes. This is in agreement with the 
possibility of choosing normal coordinates \cite{grk}, where 
such terms are absent.}.

\section{Low-energy limit}

The lagrangian of eq.~(\ref{lag}) would allow the computation of
the cross-section for $e^+e^-\to\psis\psis\gamma$ in a
generic kinematical regime. Here we restrict ourselves to the
case of a heavy superpartner spectrum, with photino, selectrons
and the scalars $S$ and $P$ beyond the production threshold at
some chosen value of $\sqrt{s}$. In this case, it is convenient
to build an effective lagrangian for the light fields by integrating
out the heavy particles. The crucial property of such an effective
lagrangian will be its dependence only on the gauge coupling $g$
and on the supersymmetry-breaking scale $|F|^{1/2}$, without
any further reference to the supersymmetry-breaking masses $(m_{S,
P}^2 \, , m^2, M)$. By standard techniques, and focussing only 
on the terms relevant for our calculation, we obtain:
\be
{{\cal L}}_{eff} =
- {1\over 4} F_{\mu\nu} F^{\mu\nu}
+ i \psieb\Dslash\psie  +{i\over 2} \psisb\dslash\psis
+\sum_{i=1}^4 {\cal O}_i \, .
\label{leff}
\ee
\begin{figure}[ht]
\vspace{-0.01cm}
\epsfig{figure=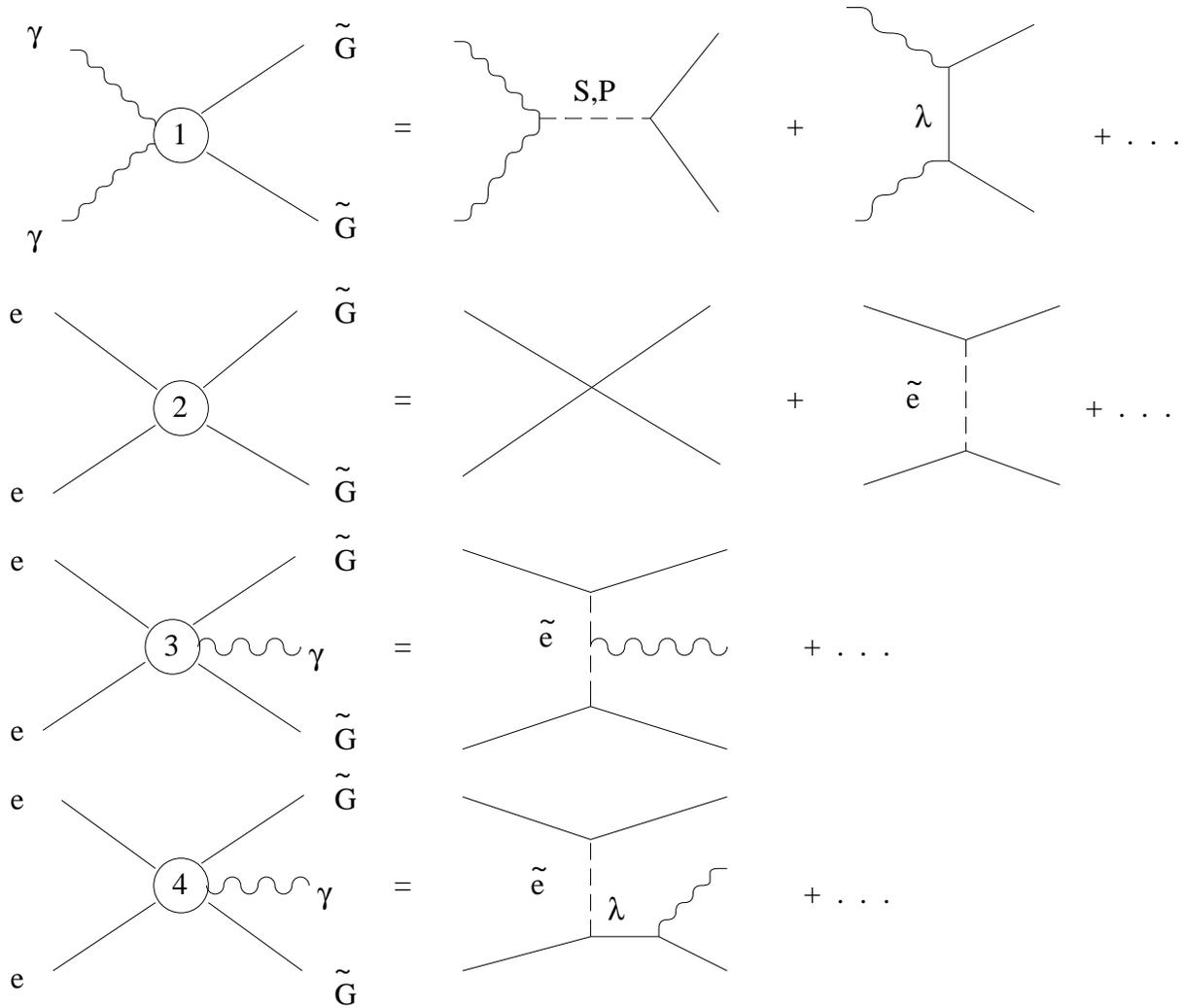,height=16cm,angle=-90}
\vspace*{0.5cm}
\caption{{\it Diagrammatic origin of the effective
operators $\co_{1 \! - \! 4}$.}}
\label{origin}
\end{figure}
The terms $\co_i$ $(i=1,...4)$ are local operators obtained through
the exchange of massive particles in the large mass limit, as shown
in fig.~\ref{origin}. Their field-dependent parts have mass dimension
$d=8$, and the corresponding coefficients scale as $1/|F|^2$. The 
operator $\co_1$ involves two photons and two goldstinos, and is 
generated \cite{bfz3} by integrating over $S$, $P$ and $\psil$:
\be
{\co_1} = - {i \over 64 |F|^2} \psisb
\left[ \gamma^\mu , \gamma^\nu \right]
F_{\mu \nu} \dslash 
\left[ \gamma^\alpha , \gamma^\beta \right]
F_{\alpha \beta} \, \psis \, .
\label{elleuno}
\ee
Notice that the form of $\co_1$ is the result of a crucial 
cancellation among individual contributions, corresponding
to $d=7$ operators with coefficients scaling as $M/F^2$.
The operator $\co_2$ is a four-fermion interaction term
among electrons and goldstinos recovered by combining the
contact term in the last line of eq.~(\ref{lag}) with
contributions originating from selectron exchanges~\cite{bfzeegg}:
\be
{\co_2} = - {1 \over 2 |F|^2} \left\{ \psieb
\psis \Box \left( \psisb \psie \right) -
\psieb \gamma_5 \psis \Box \left( \psisb
\gamma_5 \psie \right) \right\} \, .
\label{elledue}
\ee
Also in this case, the form of $\co_2$ is the result of a crucial 
cancellation among individual $d=6$ operators scaling as $m^2/F^2$. 
The operator $\co_3$ is generated by attaching a photon to a 
selectron exchanged among electrons and goldstinos. Notice that, 
despite the presence of two selectron propagators, this contribution 
does not vanish in the large $m$ limit, due to a compensating 
factor $m^4$ from the electron-selectron-goldstino vertices:
\be
{\co_3} = {i g A^{\mu} \over |F|^2} \left\{
\psieb  \psis \partial_\mu \left( \psisb \psie
\right) - \psieb \gamma_5 \psis \partial_\mu
\left( \psisb \gamma_5 \psie \right) \right\} \, .
\label{elletre}
\ee
The operators $\co_2$ and $\co_3$ are not separately gauge
invariant. Only their sum is contained in a gauge-invariant
combination, as can be seen by recasting it in the form:
\be
{\co_2}+{\co_3}+\ldots= {1 \over 2 |F|^2} \left\{ \Dmu\left(\psieb
\psis\right) \Dmd \left( \psisb \psie \right) -\Dmu\left(
\psieb \gamma_5 \psis\right) \Dmd \left( \psisb
\gamma_5 \psie \right) \right\} \, ,
\label{gaugei}
\ee
where the dots denote terms quadratic in $A_\mu$ or
vanishing on-shell. Finally, the operator $\co_4$ is
a contact term that directly contributes (as $\co_3$)
to the scattering amplitude under investigation. It
originates from a combined selectron and photino exchange:
\bea
{\co_4} & = &  {i g \over 8 |F|^2} \left\{
\psisb  \left[ \gamma^\mu , \gamma^\nu \right]
F_{\mu \nu} \left( \psie \psieb \psis - \gamma_5
\psie \psieb \gamma_5 \psis \right) \right.
\nonumber \\ & & \phantom{i e Q_e \over 8 |F|^2}
\left. + \left( \psisb \psie \psieb - \psisb
\gamma_5 \psie \psieb \gamma_5 \right)
F_{\mu \nu} \left[ \gamma^\mu , \gamma^\nu \right]
\psis \right\} \, .
\label{ellequa}
\eea
Notice that, despite the presence of a selectron and a 
photino propagator, this contribution does not vanish in the 
limit of large $m$ and $M$, due to the compensating factors 
$m^2$ and $M$ coming from the electron-selectron-goldstino 
and the photon-photino-goldstino vertices, respectively.
\begin{figure}[ht]
\vspace{-0.01cm}
\begin{center}
\epsfig{figure=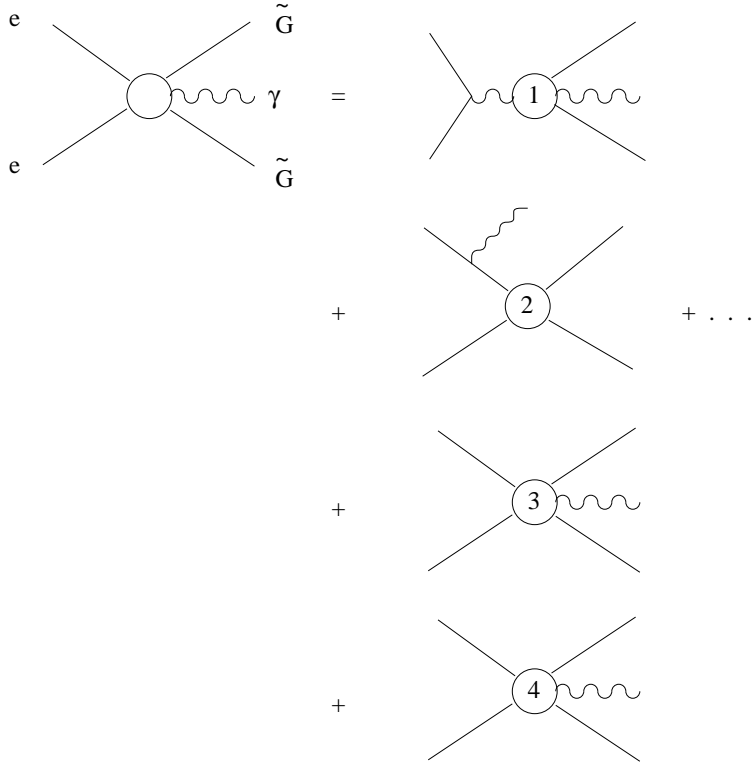,height=10cm,angle=0}
\end{center}
\vspace*{0.5cm}
\caption{{\it Feynman diagrams contributing to $e^+ e^- \to \grav
\grav \gamma$ in the low-energy effective theory.}}
\label{diagrams}
\end{figure}

It is now possible to list the independent amplitudes contributing,
in the low-energy regime, to the scattering process $e^+e^-\to\tg\tg
\gamma$. They are shown in fig.~\ref{diagrams}. The contribution of
the operator $\co_1$  is obtained by attaching an electron-positron
line to one of the photons. The operator $\co_2$ contributes via
initial state radiation, and the remaining operators are by
themselves local contributions to the process. The total amplitude 
is therefore the sum of four terms:
\be
\label{emmetot}
\cm = \cm_1 + \cm_2 + \cm_3 + \cm_4 \, ,
\ee
with the following expressions in momentum space:
\be
\label{emme1}
\cm_1 =
{i g \over 16 |F|^2} {1 \over (p_1 + p_2)^2}
\left[ \vbd \gmd \upu \right]
\left[ \usbu {{\cal A}}_1^{\mu} \vsd
-
\usbd {{\cal A}}_2^{\mu} \vsu \right]
\, ,
\ee
\bea
\cm_2 & = &
- {i g \over |F|^2} \vbd \left\{
\esl {1 \over \ksl-\pdsl} {{\cal A}}_3
+ {{\cal A}}_4 {1 \over \pusl - \ksl} \esl
\right. \nonumber \\ & & \left.
- \esl {1 \over \ksl-\pdsl} \gc {{\cal A}}_3 \gc
- \gc {{\cal A}}_4 \gc {1 \over \pusl - \ksl} \esl
\right\} \upu \, ,
\label{emme2}
\eea
\be
\cm_3 =
- {i g \over |F|^2} \emu \vbd \left\{
{{\cal A}}_5^{\mu} - \gc {{\cal A}}_5^{\mu} \gc \right\}
\upu \, ,
\label{emme3}
\ee
\be
\cm_4 =
{i g \over 4 |F|^2} \vbd \left\{
{{\cal A}}_6 - \gc {{\cal A}}_6 \gc \right\}
\upu \, ,
\label{emme4}
\ee
where
\be
{{\cal A}}_1 =
[\pusl+\pdsl,\gmu] (\ksl+\qdsl) [\ksl,\esl]
-
[\ksl,\esl] (\ksl+\qusl) [\pusl+\pdsl,\gmu]
\, ,
\label{auno}
\ee
\be
{{\cal A}}_2 =
[\pusl+\pdsl,\gmu] (\ksl+\qusl) [\ksl,\esl]
-
[\ksl,\esl] (\ksl+\qdsl) [\pusl+\pdsl,\gmu]
\, ,
\label{adue}
\ee
\be
{{\cal A}}_3 =
q_1 \cdot p_1 \vsd \usbu
-
q_2 \cdot p_1 \vsu \usbd
\, ,
\label{atre}
\ee
\be
{{\cal A}}_4 =
q_2 \cdot p_2 \vsd \usbu
-
q_1 \cdot p_2 \vsu \usbd
\, ,
\label{aqua}
\ee
\be
{{\cal A}}_5^{\mu} =
(p_1-q_1)^{\mu} \vsd \usbu
-
(p_1-q_2)^{\mu} \vsu \usbd
\, ,
\label{acin}
\ee
\be
{{\cal A}}_6 =
[\vsd \usbu - \vsu \usbd]
[\ksl,\esl]
+
[\ksl,\esl]
[\vsd \usbu - \vsu \usbd]
\, .
\label{asei}
\ee
We have denoted by $p_1$, $p_2$, $q_1$, $q_2$ and $k$ the four-momenta
of the incoming electron and positron and of the outgoing goldstinos
and photon, respectively. Notice that, due to the structure of the
fermionic currents in eqs.~(\ref{emme1})--(\ref{emme4}) and to the
Majorana nature of the goldstino, the only non-vanishing amplitudes
are those where electron and positron have opposite helicities, and
the two goldstinos have opposite helicities. Taking into account the
two possible photon polarizations, we conclude that eight out
of sixteen helicity configurations do not contribute to the process.

\section{Cross-section for $e^+ e^- \to \grav \grav \gamma$}

Starting from the amplitudes of eqs.~(\ref{emmetot})--(\ref{asei}), 
we computed the 
double differential cross-section $d \sigma/(d \xg d  \ctg)$, where
$\xg$ is the fraction of the beam energy carried by the photon, and
$\theta_\gamma$ is the scattering angle of the photon with respect
to the direction of the incoming electron, in the centre-of-mass
frame. To cross-check our computation, we used two independent
methods. One method proceeded through the evaluation of the helicity 
amplitudes in the centre-of-mass frame, with explicit expression 
for spinors, four-momenta and polarization vectors. The second
method consisted in the direct computation of the unpolarized
cross-section, by means of standard trace techniques, with the help
of the program Tracer \cite{tracer}. After integrating over part
of the phase space, we obtained the following result:
\be
{d^{\, 2} \sigma \over d \xg d \cos \theta_{\gamma}}
=
{\alpha s^3 \over 320 \pi^2 |F|^4}
\cdot f ( \xg, \ctg) \, ,
\label{dsdxdct}
\ee
where
\be
f ( \xg, \ctg)  = 2 (1 - \xg)^2
\left[ {(1 - \xg) (2 - 2 \xg + \xg^2) \over \xg \sqtg}
+ 
{\xg (-6 + 6 \xg + \xg^2) \over 16}
-
{\xg^3 \sqtg \over 32} \right] \, .
\label{fxct}
\ee
\begin{figure}[ht]
\vspace{-0.1cm}
\epsfig{figure=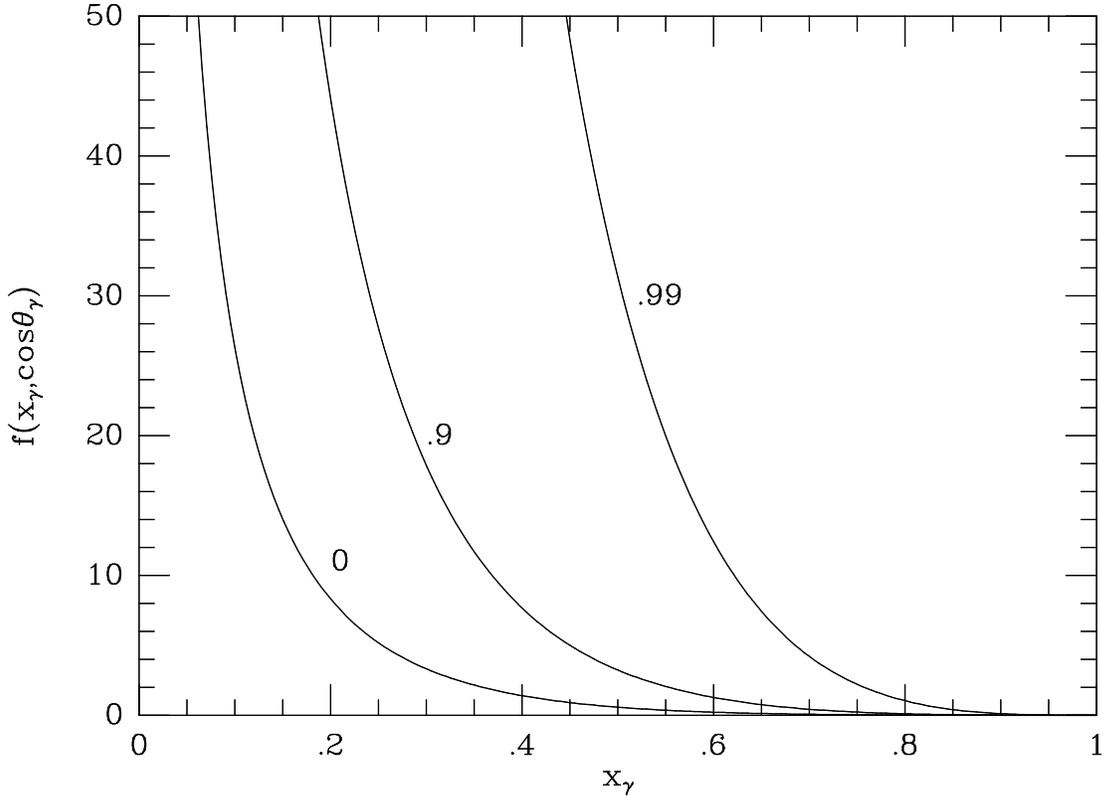,height=6cm,angle=-90}
\vspace*{3.0cm}
\caption{{\it The function $f$ of eq.~(\ref{fxct}), plotted versus
$\xg$, for the indicated representative values of $|\ctg|$.}}
\label{xgdep}
\end{figure}
\begin{figure}[ht]
\vspace{-0.1cm}
\epsfig{figure=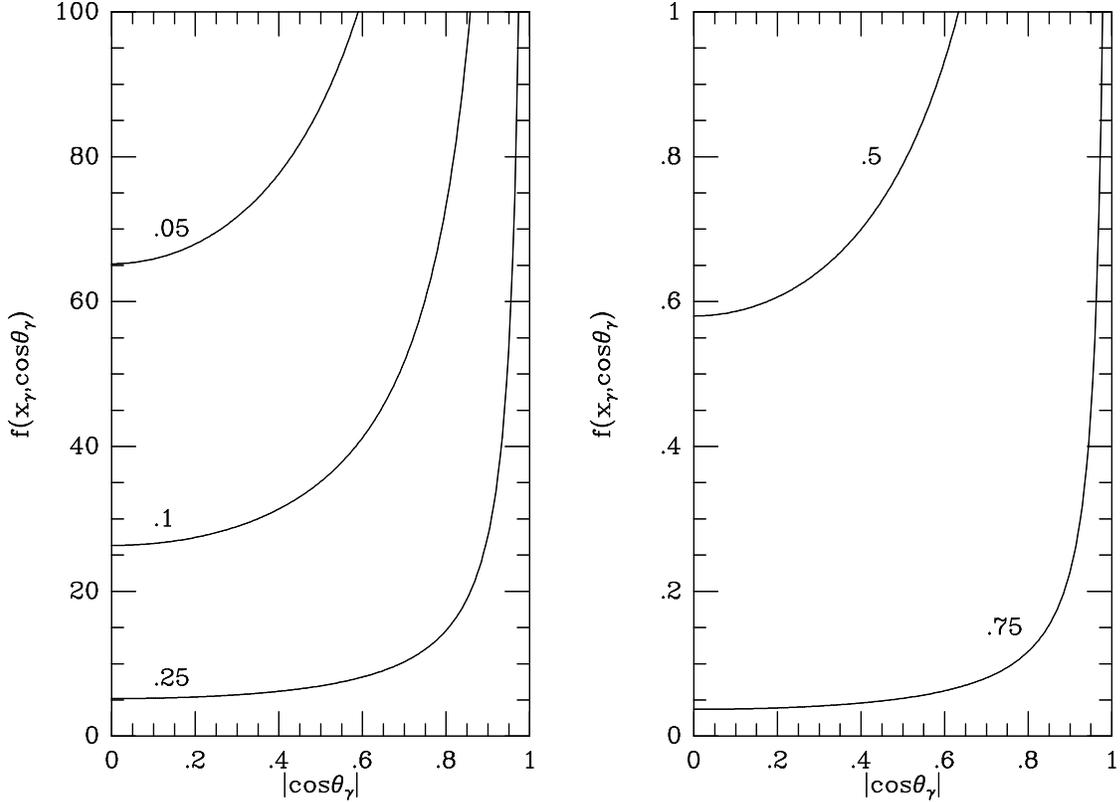,height=6cm,angle=-90}
\vspace*{3.0cm}
\caption{{\it The function $f$ of eq.~(\ref{fxct}), plotted versus
$|\ctg|$, for the indicated representative values of $\xg$ (notice
the different scale in the two windows).}}
\label{tgdep}
\end{figure}
\begin{figure}[ht]
\vspace{-0.1cm}
\epsfig{figure=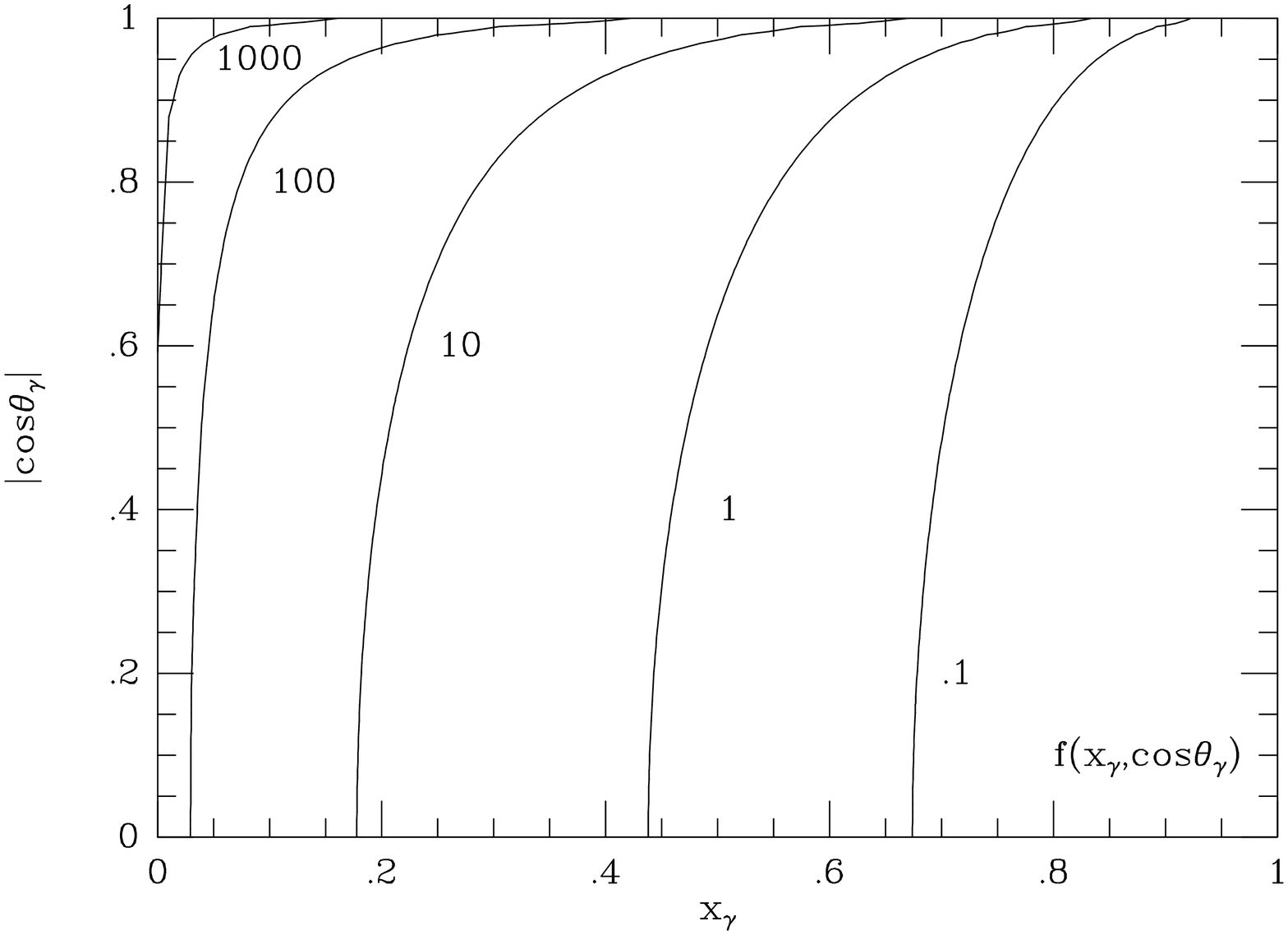,height=6cm,angle=-90}
\vspace*{3.0cm}
\caption{{\it Contours of the function $f$ of eq.~(\ref{fxct})
in the $(\xg,|\ctg|)$-plane.}}
\label{fcont}
\end{figure}
Notice that the dependence on $\sqrt{s}$ and on the
supersymmetry-breaking scale $|F|^{1/2}$ can be factorized
out of the function $f$, which carries instead all the dependence
on the photon fractional energy $\xg$ and on the photon
angle $\theta_\gamma$. For these reasons, it is convenient
to illustrate our results in terms of the function $f$.
Figs.~\ref{xgdep} and \ref{tgdep} show the dependence of
$f(\xg,\ctg)$ on each of its two variables, keeping the other
one fixed to some representative value. Contours corresponding
to constant values of $f$ in the $(\xg,|\ctg|)$ plane are
displayed in fig.~\ref{fcont}. 

It is interesting to observe that, for the soft/collinear
part of the photon spectrum, $\xg \ll 1$ and/or $\sin^2
\theta_{\gamma} \ll 1$, our result could have been derived 
by considering the total cross-section $\sigma_0$ for the 
process $e^+ e^- \to \grav \grav$ and applying a standard 
approximate formula for photon radiation \cite{soft}:
\be
\label{appro}
{d \sigma \over d \xg d \cos \theta_{\gamma}} \simeq
\sigma_0 [ \hat{s} ] \cdot {\alpha \over \pi}
{1 + (1 - \xg)^2 \over \xg \sqtg}  \, ,
\ee
where $\sigma_0$ is evaluated for $\hat{s} \equiv (1 - \xg)
s$. Computing $\sigma_0$ directly from the effective operator of 
eq.~(\ref{elledue}), we found~\cite{bfzeegg}:
\be
\label{seegg}
\sigma_0 \equiv \sigma (e^+ e^- \to \grav \grav) =
{s^3 \over 160 \pi |F|^4} \, .
\ee
Plugging this result into (\ref{appro}) does indeed
reproduce the leading term of the above result
(\ref{dsdxdct})--(\ref{fxct}).
As expected, for relatively soft/collinear photons the
cross-section is dominated by initial-state
radiation, and the signal must compete with
the irreducible Standard Model background coming
from $e^+ e^- \to \nu \ov{\nu} \gamma$. This
suggests that, in the absence of an anomaly
with respect to the Standard Model predictions
and applying rather loose cuts on the photon
spectrum, the approximation of eq.~(\ref{appro})
should be sufficient to obtain a lower
bound on the gravitino mass. On the other hand,
in the presence of an anomaly the specific
features of the spectrum may be useful in
establishing the significance of its
interpretation in terms of a superlight
gravitino. 

From the double differential cross-section of
eqs.~(\ref{dsdxdct}) and (\ref{fxct}), the total
cross-section can be easily computed. We recall that,
because of the QED infrared and collinear singularities,
we need to specify some cuts on the photon spectrum. To
illustrate our results, we specify our cuts in
terms of a minimum value for $\xg$ and a maximum
value for $|\ctg|$, to be denoted by $\xgmin$ and
$\ctmax$, respectively\footnote{Alternatively, we
could have chosen to cut in $(\xg)_T \equiv \xg \,
\sin \theta_\gamma$ rather than in $\xg$.}. The
integrated cross-section is then given by
\be
\sigma = {\alpha s^3 \over 320 \pi^2 |F|^4}
\cdot I ( \xgmin \, , \ctmax) \, ,
\label{intcs}
\ee
where the integral
\be
I ( \xgmin, \ctmax) = \int_{\xgmin}^1 \!\!\! d \xg
\int^{\ctmax}_{- \ctmax} \! d \ctg \; f ( \xg, \ctg)
\label{int}
\ee
can be evaluated analytically by elementary methods.
Table~\ref{sigint} gives the values of the integral $I$,
defined in eq.~(\ref{int}), for some representative
choices of the cuts $\xgmin$ and $\ctmax$.
\begin{table}[htb]
$$
\begin{array}{|cc|c|c|c|c|c|c|}
\hline
& & & & & & & \\
& \ctmax & 0.7 & 0.8 & 0.85 & 0.9 & 0.95 & 0.975 \\
\xgmin & & & & & & & \\
& & & & & & & \\
\hline
& & & & & & & \\
0.2 & & 1.59 & 2.02 & 2.31 & 2.72 & 3.39 & 4.05 \\
& & & & & & & \\
\hline
& & & & & & & \\
0.15 & & 2.54 & 3.22 & 3.68 & 4.32 & 5.39 & 6.44 \\
& & & & & & & \\
\hline
& & & & & & & \\
0.1 & & 4.21 & 5.34 & 6.11 & 7.17 & 8.93 & 10.7 \\
& & & & & & & \\
\hline
& & & & & & & \\
0.05 & & 7.79 & 9.87 & 11.3 & 13.2 & 16.5 & 19.7 \\
& & & & & & & \\
\hline
\end{array}
$$
\caption{{\it Values of the integral $I$, defined in eq.~(\ref{int}),
for some representative choices of the cuts $\xgmin$ and $\ctmax$.}}
\label{sigint}
\end{table}

The process under consideration can give rise to an observable
signal at LEP, characterized by a single photon plus missing 
energy. From eq.~(\ref{intcs}), we see that an experimental upper 
bound $\sigma_{exp}$ on the signal cross-section, obtained at the energy 
$\sqrt{s}$ and for a given set of cuts, corresponding to a definite 
value of the integral $I$, translates into a lower limit on the 
supersymmetry breaking scale $|F|^{1\over 2}$:
\be
|F|^{1\over 2} > 125~{\rm GeV}
\left[{\sqrt{s}~({\rm GeV})\over 200}\right]^{3\over 4}
\left[{I\over \sigma_{exp}~({\rm pb})}\right]^{1\over 8}~~,
\label{blep2}
\ee
or, in terms of the gravitino mass and remembering 
eq.~(\ref{flat}):
\be
m_{3/2} > 3.8\cdot 10^{-6}~{\rm eV}
\left[{\sqrt{s}~({\rm GeV})\over 200}\right]^{3\over 2}
\left[{I\over \sigma_{exp}~({\rm pb})}\right]^{1\over 4}~~.
\label{bmlep2}
\ee
Notice that these limits are mainly sensitive to the energy
$\sqrt{s}$, while the dependence on $\sigma_{exp}$ is quite mild.
For example, a factor 10 reduction in $\sigma_{exp}$ would increase
the lower bounds on $|F|^{1\over 2}$ and $m_{3/2}$ by a factor 
1.3 and 1.8, respectively. Similarly, the limits do not depend 
strongly on the precise choice of the cuts, within the set 
displayed in table~1. Up to now, the highest energy reached by 
LEP is 183 GeV, where each of the four experiments has collected 
an integrated luminosity of about 60 pb$^{-1}$. Both at $\sqrt{s} 
= 183 \gev$ and at lower energies, no anomalies have been reported 
so far \cite{lep2} in the photon plus missing energy channel, 
which, in the Standard Model, is dominated\footnote{At small 
angle, one should also consider the reducible background 
due to $e^+ e^- \to (e^+ e^-) \, \gamma$ and $e^+ e^- \to
(\gamma \gamma) \, \gamma$, where the particles in brackets are
undetected.} by the process $e^+e^-\to \nu{\bar\nu}\gamma$. 
We leave to our experimental colleagues the detailed analysis of 
the resulting bounds. In the meantime, we can tentatively assume 
an upper bound of $0.2$~pb on the signal cross-section $\sigma_{exp}$, 
integrated over $|x_\gamma| > 0.05$ and $|\cos\theta_\gamma| < 0.95$.
Taking into account that, for our representative cuts, the Standard 
Model cross-section for $e^+e^-\to \nu{\bar\nu}\gamma$ is 
approximately 5~pb, with a pronounced peak of the photon spectrum 
around the `radiative return to the $Z$', we obtain the lower bound
\be
|F|^{1\over 2} > 200 \gev \, ,
\ee
corresponding to the following lower bound on the
gravitino mass:
\be
m_{3/2} > 10^{-5} \ev \, .
\ee

\section{Discussion}

As a first comment, we would like to remark that, even if our
results were obtained in a simplified model, they remain valid 
when we include the full gauge group and matter content of the 
MSSM, as long as all supersymmetric particles but the gravitino 
cannot be produced on-shell.

We also recall that our computation was based on the explicit
integration of the heavy superpartners in the low-energy limit,
starting from a generic theory where supersymmetry is linearly
realized but spontaneously broken by an $F$-term, with negligible
Fayet-Iliopoulos and higher-derivative terms. This led to the 
non-linear realization of global supersymmetry associated with 
the effective lagrangian (\ref{leff}), and finally to
explicit expressions for the differential and integrated 
cross-sections. However, as recently discussed in \cite{bfzeegg},
such non-linear realization is not unique. For example, the 
authors of \cite{nach} computed the same cross-sections in
a different non-linear realization \cite{wess} (believed 
then to be unique), and found a different result, although
with the same dependence on $F$ and $s$ as ours. To relax
the assumptions of the present calculation, and provide a
framework encompassing the present case, the case of \cite{nach}
and additional possibilities, we would need a general 
parametrization of the possible non-linear realizations, at 
least for the terms that affect the process under consideration. 
Unfortunately, such a general formulation is not yet available. 
Neglecting the electron mass and the mixing in the selectron 
sector, however, we know from \cite{bfzeegg} the most general 
form of the local four-fermion effective interaction that replaces 
$\co_2$ [see eq.~(\ref{elledue})] in an arbitrary non-linear 
realization. It depends on two free parameters, $\alpha_L$ and
$\alpha_R$, associated with the contributions of $e_L \equiv 
P_L e$ and $e_R \equiv P_R e$, respectively. Taking advantage 
of the fact that, for realistic experimental situations, the 
bulk of the cross-section is well reproduced by the approximate 
formula of eq.~(\ref{appro}), we give here \cite{bfzeegg} the 
corresponding general unpolarized cross-section $\sigma_0^{GEN}$:
\be
\sigma_0^{GEN} ( e^+ e^- \to \grav \grav) = 
{s^3 \over 15360 \pi |F|^4} \left[ 
(8 + 10 \alpha_L + 5 \alpha_L^2)
+
(8 + 10 \alpha_R + 5 \alpha_R^2)
\right] \, .
\ee
The results of the present paper are then obtained, in the approximation 
of eq.~(\ref{appro}), for $\alpha_L=\alpha_R=-4$, those of \cite{nach} 
for $\alpha_L = \alpha_R =0$. The minimum value of the cross-section is 
obtained for $\alpha_L = \alpha_R = -1$:
\be
\label{smin}
\sigma_0^{min} ( e^+ e^- \to \grav \grav) = 
{s^3 \over 2560 \pi |F|^4} \, .
\ee
In the absence of experimental anomalies, the combination of
eq.~(\ref{smin}) and eq.~(\ref{appro}) can be safely used to 
establish a model-independent lower bound on the gravitino mass. 
Because of the strong and universal power-law behaviour of the 
cross-section, always proportional to $s^3/|F|^4$, this bound 
is rather stable with respect to variations of the parameters 
$\alpha_L$ and $\alpha_R$ over plausible ranges. However, should 
a signal show up at LEP or at future linear colliders, having the 
full expression of the cross-section for the most general class 
of models would be very important, since a detailed analysis of 
the photon spectrum would offer the unique opportunity of 
distinguishing among possible fundamental theories.

In the present paper, we have assumed that all exotic
particles besides the gravitino are far from the
production threshold at the available energy, so 
that the local operators $\co_1$--$\co_4$ provide
a good approximation of the dynamics. An advantage
of our approach, with respect to the use of non-linear 
lagrangians, is that, by relaxing the kinematical assumption
about superparticle masses, we could take into account
also propagator effects, which might be important if a 
signal were detected. 

Concerning the energy dependence of the cross-sections
analyzed here, we recall that the authors of \cite{lp} 
have recently proposed an interaction term among two 
goldstinos and a photon that reads:
\be
\delta{\cal L}={\mu^2\over F^2} \left( \partial^\mu
\ov{\grav} \right) \gamma^\nu\grav F_{\mu\nu}~~,
\label{lpterm}
\ee
where $\mu^2$ is an independent mass parameter. If
present, and in the absence of cancellation mechanisms,
such a term would induce a local four-fermion operator,
involving two goldstinos and an electron-positron pair,
with dimension $d=6$, and characterized by a dimensionful 
coupling ${\cal O} (\mu^2/|F|^2)$. Such a four-fermion operator 
would contribute to the cross-section for $e^+ e^- \to \grav 
\grav$ with terms scaling as $s \mu^4 / |F|^4$. As a consequence, 
also the scaling properties of the cross-section for $e^+ e^- 
\to \grav \grav \gamma$ would be modified. Depending on the 
value of $\mu^2$ and on the typical energy of the process, 
the gravitino phenomenology could be more or less heavily 
affected. However, the term of eq.~(\ref{lpterm}) does not 
arise in our calculational framework. More importantly, it was 
shown in full generality \cite{bfzeegg} that local four-fermion
operators such as those mentioned above are not allowed by 
supersymmetry. This leads us to the belief that the energy 
dependence of the cross-section derived here will survive in 
the most general case.

The processes considered here, or other processes controlled by 
the same effective interactions, could in principle be relevant 
for nucleosynthesis or stellar cooling \cite{ascos,bfz3,lp}. In 
general, however, the bounds coming from high-energy colliders leave
little room for interesting effects in astrophysics and cosmology
\cite{bfz3}, since the relevant cross-sections have a strong
positive power-dependence on the energy, and the typical energies
involved are much smaller than the present collider energies.

Bounds coming from the finite goldstino-smuon loop contribution 
to the anomalous magnetic moment of the muon \cite{gmtwo} are not 
very stringent either. For smuon masses up to 1~TeV, they do not
constrain $m_{3/2}$ more strongly than the processes discussed in
this paper. Moreover, ambiguities similar to the ones discussed 
for our result should be taken into account, since the operator 
associated with $(g-2)_{\mu}$, when supersymmetrized, corresponds 
to a higher-derivative term, to which we may associate an arbitrary 
counterterm if we do not have additional informations on the 
fundamental theory.

In summary, high-energy colliders are by far the best environment
to test the possible existence of a superlight gravitino. The
present LEP data, analysed as discussed in the present paper, 
should allow to establish an absolute lower bound $m_{3/2}
\simgt 10^{-5} \ev$. The prospects for the present Tevatron
data are at least as good, since the higher available energy
should be more than enough to compensate for the more difficult
experimental environment \cite{bfmz}.
\vfill{
\section*{Acknowledgements}
We would like to thank P.~Checchia, A.~De~Min, C.~Dionisi,
S.~Giagu, P.~Nason, O.~Nicrosini and G.~Ridolfi for useful 
discussions, and A.~Vicini for advice on computational matters.}

\newpage

\end{document}